\documentclass[twocolumn,showpacs,preprintnumbers,amsmath,amssymb]{revtex4}

\usepackage{epsfig}
\usepackage{graphics}
\usepackage{graphicx}
\usepackage{dcolumn}
\usepackage{bm}

\begin{document}


\title{Delay enhanced coherent chaotic oscillations in networks with
large disorders }
\author{D.~V.~Senthilkumar$^{1}$}
\author{R.~Suresh$^2$}
\author{Jane H.~Sheeba$^2$}
\author{M.~Lakshmanan$^2$}
\author{J.~Kurths$^{1,3}$}

\affiliation{
$^1$Potsdam Institute for Climate Impact Research, 14473 Potsdam, Germany\\
$^2$Centre for Nonlinear Dynamics, Department of Physics, Bharathidasan University, Tiruchirapalli 620 024, India\\
$^3$Institute of Physics, Humboldt University, 12489 Berlin, Germany}
\date{\today}

\begin{abstract}  
We study the effect of coupling delay in a regular network with a ring topology 
and in a more complex network with an all-to-all (global) topology in the presence of 
impurities (disorder). We find that the coupling delay is capable of inducing phase
coherent chaotic oscillations in both types of networks thereby enhancing the spatiotemporal
complexity even in the presence of $50\%$ of symmetric disorders of both fixed and random 
types.  Further,
the coupling delay increases the robustness of the networks upto $70\%$ of disorders,
thereby preventing the network from acquiring periodic oscillations to foster disorder-induced 
spatiotemporal order. 
We also confirm the enhancement of coherent chaotic oscillations using 
snapshots of the phases and values of the associated Kuramoto order parameter.
We also explain a possible mechanism for the phenomenon
of delay-induced coherent chaotic oscillations despite the presence of
large disorders and discuss its applications.
\end{abstract}

\maketitle

\section{\label{sec:level1}Introduction}
In recent times, researchers have been interested in studying networks of oscillators 
with time-delayed  coupling because of their wide applications in 
science \cite{schuster1989,niebur1991,kozyreff2000,wirkus2002}, 
engineering and technology \cite{heath2000,lindsey1996,reddy2000}. 
Considering the fact that in most realistic physical and biological 
systems \cite{glass1977,cooke1982,wischert1994} the interaction signal 
propagates through media with limited speed, its finite signal propagation 
time induces a time-delay in the received signal \cite{hu2001,dvsbook,swadlow1985}. 
For example, in biological neural networks, it has been shown 
that the neural connections are full of variable loops such that the propagation 
of signal through the loops can result in a large time-delay (synaptic delay), 
and it is also reported that the axons can generate time-delay upto 
$300~ms$~\cite{swadlow1985}.  A typical nonlinear time-delay system is 
a veritable black box~\cite{dvsbook}  and that the
delay coupling itself gives rise to a plethora of novel phenomena, such as delay-induced 
amplitude death \cite{reddy2000}, phase-flip bifurcation \cite{prasad2006}, 
synchronizations of different types \cite{dhamala2004}, 
multistability \cite{shayer2000}, chimera states \cite {sethia2008,jane2008}, etc.
in coupled nonlinear oscillator systems.
\begin{figure}
\centering
\includegraphics[width=0.7\columnwidth]{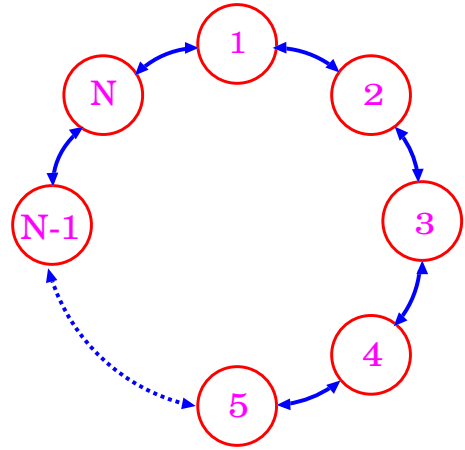}
\caption{\label{fig1} (Colour online) The schematic diagram of the array of pendula with periodic 
boundary conditions.}
\end{figure}

In this paper, we consider a network of forced and damped nonlinear pendula 
studied earlier by Braiman et al.\cite{braiman1995}, commonly known as the 
forced Frenkal-Kontorova model. It represents a straightforward physical
realization of an array of diffusively coupled Josephson junctions 
\cite{ustinov1993,pagano1986} in which the applied current in each junction 
is modulated by a common frequency. The possibility of obtaining synchronized 
motion in one and two-dimensional chaotic arrays of  such systems has been 
investigated in Refs. \cite{weiss2001,agtk1998,zhou2008,gavrielides1998}, 
where the complex chaotic behavior of the collective systems  was completely tamed when a 
certain amount of  impurities (disorder) was introduced.
To be specific, disorder enhanced spatiotemporal 
regularity \cite{braiman1995,qi2003}, disorder enhanced synchronization 
\cite{brandt2006,braimanditto1995} and taming chaos with disorder 
\cite{weiss2001,agtk1998,gavrielides1998} in such systems have received central importance in  
recent research on complex systems and their applications.

In our present studies, we study a regular network with a ring topology and
a more complex network with an all-to-all (global) topology with different
densities (sizes) of impurities (disorder) and examine the
effect of time delay in the coupling.
In particular, the oscillations of each pendulum affects the oscillations of the pendula to 
which it is connected to, after some finite time-delay $\tau$. 
In such a configuration, we are interested in investigating the possibility of
achieving coherent chaotic dynamics in the network despite of the presence
of a substantial amount of disorder and, thereby, enhancing the spatiotemporal
complexity, a counter-intuitive result to the expected (reported) outcome of taming chaos
and enhancing spatiotemporal order with even a negligible size of disorder
in the network (in the absence of coupling delay).  Here by coherent chaotic dynamics, we
mean the emergence of collective (phase-coherent or phase synchronized) chaotic 
oscillations (but not complete synchronization) in the entire network
despite the presence of disorder~\cite{expone}. 
The delay enhanced phase-coherent chaotic oscillations are characterized  
both qualitatively and quantitatively
using snapshots of the phases of the pendula in the networks and 
the Kuramoto order parameter~\cite{yk1984}. Recently, similar 
coherent states have been observed in Bose-Einstein
condensates on tilted lattices for strong field showing highly organized patterns,
often denoted as quantum carpets~\cite{arkeag2010}. Enhancing spatiotemporal complexity
or at least preserving the original spatiotemporal pattern in the midst of a noisy
environment due to the presence of disorder is crucial for applications,
such as spatiotemporal and/or secure communication~\cite{ojalvo2001}  
in spatially extended systems, especially in  biology and
physiology \cite{schiff1994}, in the state of art of modern computing, namely 
liquid state machines (LSM), in which
the degree of spatiotemporal complexity of the network of dynamical systems
determines the highest degree of computational performance (i.e, mixing property)
\cite{maass2009}, etc.

In particular, we will show that time-delay in the coupling induces coherent
chaotic oscillations of the  network of coupled systems, in both diffusively coupled
pendula with periodic boundary conditions and in a globally coupled network, thereby enhancing the
spatiotemporal complexity despite the presence of a large number of disorders,
even upto half the size of the network. 
Further, coupling delay enhances the robustness of the network against
disorders of size greater than $50\%$ of the network thereby preserving the
original dynamical states of the network and preventing disorder-enhanced synchronous
periodic oscillations of the entire network leading to spatiotemporal order.
It is to be noted that in an array without delay even the presence of a very 
small periodic disorder itself is capable of
suppressing the chaotic oscillations of the entire network and thereby  inducing 
spatiotemporal regularity as demonstrated in Refs.~\cite{weiss2001,agtk1998,gavrielides1998}. 
We will also explain an appropriate mechanism for delay-induced coherent
chaotic oscillations leading to enhanced spatiotemporal complexity based on
a mechanism for taming chaoticity (in the absence of delay), 
as reported in~\cite{braiman1995}. A relevant study focusing on macroscopic 
properties of the globally connected heterogeneous neural network has
revealed similar irregular collective behavior~\cite{sfap2010}.

The paper is organized as follows. In Sec.~\ref{sec:nd}, we will briefly discuss the 
existing results on taming chaoticity leading to spatiotemporal regularity 
without any delay coupling for a linear array of nonlinear pendula with periodic 
boundary conditions, which will be useful for a later comparison.  
We will demonstrate our results on delay-induced
coherent chaotic oscillations despite the presence of large disorders, even upto $70\%$,
in Sec.~\ref{sec:level2}. Similar results are presented in a network of
globally coupled pendula both with and without delay coupling in Sec.~\ref{sec:level3}.
Finally, in Sec.~\ref{sec:level4}, we discuss our results and conclusions.

\section{\label{sec:nd}Linear array of pendula in the absence of coupling delay}
We consider a chain of $N$ forced coupled nonlinear pendula whose
equation of motion can be written  
as~\cite{braiman1995,brandt2006,weiss2001,agtk1998,gavrielides1998,braimanditto1995,qi2003}
\begin{subequations}
\begin{eqnarray}
ml^2\dot{x_i} & = & y_i,\\
\dot{y_i} & = & - \gamma y_{i}-mgl~sin~x_{i}+f+f^{\prime}_i sin(\omega t)+ \nonumber\\
& &C[y_{i+1}(t)-2y_{i}(t)+y_{i-1}(t)],
\end{eqnarray}
\label{nodelay_eqn1}
\end{subequations}
where $i = 1,2,\cdots,N$. We choose the following periodic boundary conditions: 
$x_0=x_N$ and $x_{N+1}=x_1$. The parameters are taken as follows: the 
mass of the bob $m = 1.0$, the damping $\gamma = 0.5$, acceleration due to the
gravity $g = 1.0$, dc torque $f = 0.5$, angular frequency $\omega = 0.67$, pendulum 
length $l = 1.0$, $f^{\prime}_i=f^{\prime}$ is the ac torque 
and $C$ is the coupling strength.  The schematic diagram of the coupling configuration is 
shown in Fig.~\ref{fig1}, in which the first pendulum is 
connected with the second and the $N^{th}$ pendulum so that each pendulum 
gets two inputs, without any delay, from its nearest pendula. For the coupling strength $C = 0.0$, 
the pendula are uncoupled and evolve independently. 

\begin{figure*}
\centering
\includegraphics[width=1.8\columnwidth]{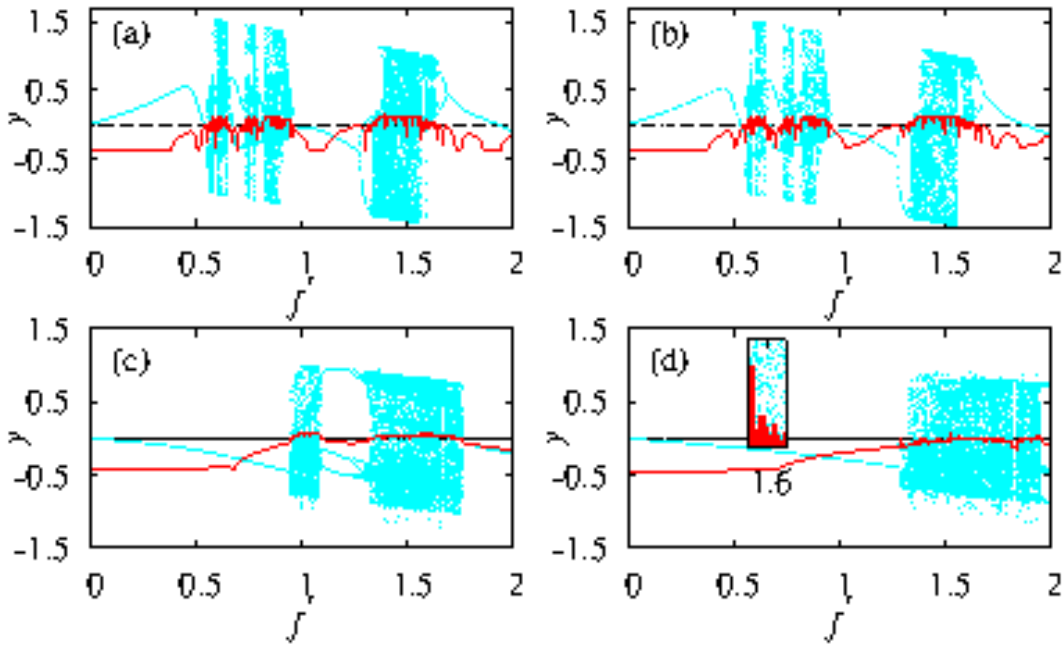}
\caption{\label{bif}(Color online) Bifurcation diagrams of a single pendulum in 
a ring of three coupled pendula and the largest Lyapunov exponent (of a single pendulum for $C=0.0$
and that of the entire network for $C>0.0$)
for different values of the coupling strength $C$ and the coupling delay $\tau$.
(a) $C=0.0$ and $\tau=0.0$,
(b) $C=0.5$ and $\tau=0.0$,
(c) $C=0.5$ and $\tau=1.5$, and
(d) $C=0.6$ and $\tau=2.0$ (inset shows that the pendulum exhibits chaotic oscillations
for $f^\prime_d=1.6$).Red (dark grey) line corresponds to the largest Lyapunov exponent and
light blue (light grey) dots correspond to the bifurcation diagram. Note that in all the
cases (b)-(d) the three pendula are in a completely synchronized state and, hence, the largest 
Lyapunov exponent corresponds to the synchronization manifold.}
\end{figure*}
In earlier studies~\cite{braiman1995,brandt2006,weiss2001,agtk1998,gavrielides1998,braimanditto1995,qi2003}, 
the authors have dealt with an array of pendula with diffusive coupling but 
without delay and have shown that the chaotic dynamics of the array is 
controlled by the inclusion of impurities, which are disorders in their 
natural frequencies and/or distributed initial phases of the external forces. In particular, 
in Ref.~\cite{weiss2001,agtk1998,gavrielides1998}, the authors have considered 
a chain of diffusively coupled 
pendula without delay and have shown that inclusion of even a single
periodic impurity is enough to tame chaos in a long chain of length 
with $N = 512$. However, we would like to point out that we are not able
obtain the results with a single impurity as reported by these authors.
Nevertheless, taming chaos and achieving spatiotemporal regularity can be
obtained for $20\%$ of impurities for appropriate coupling strength for different sizes of the array as 
reported by other authors~\cite{braiman1995,braimanditto1995,qi2003,brandt2006}.
In the following, we will briefly illustrate the results of taming chaos and achieving
spatiotemporal regularity in an array of $N=50$ coupled pendula, 
Eq.~(\ref{nodelay_eqn1}), with ring configuration
without any coupling delay to appreciate the effect of delay coupling
in the following sections. The results have been confirmed for the case of $N=512$ too.

We introduce disorder in the network of chaotic pendula by allowing one or more 
pendula to oscillate periodically as in the earlier reports~\cite{braiman1995,braimanditto1995,qi2003,brandt2006}.  
In order to fix the parameters 
(of the pendula) corresponding to the chaotic and periodic regions,
we start our analysis by plotting the bifurcation diagram of a single pendulum 
as a function of the ac torque in the range $f^{\prime}\in(0,2)$ for fixed values
of the other parameters. The bifurcation diagram and its corresponding largest 
Lyapunov exponent is depicted in Fig.~\ref{bif}(a), which
exhibit a typical bifurcation scenario leading to chaotic behavior for appropriate values of the ac torque.
To elucidate the dynamical behavior of the ring of $N$ coupled pendula as a function of a parameter,
we have explored an array of $N=3$ pendula with periodic boundary conditions in plotting the 
bifurcation diagram, because each pendulum in an array of arbitrary length $N > 2$ is coupled with its 
nearest neighbors and so each of the pendula effectively gets two inputs from its neighbors. 
Therefore the basic configuration of $N=3$ pendula is sufficient to explain the bifurcation pattern of
$N$ coupled pendula in a ring configuration for same values of the parameters. 
Indeed, we have confirmed that the bifurcation diagram remains the same irrespective of the value 
of $N$ for the same set of  parameter values as in Fig.~\ref{bif}. 
The bifurcation diagram  of a single pendulum 
in a ring of $N=3$ coupled pendula and the largest Lyapunov exponent of the
entire network for the value of the coupling strength $C=0.5$
in the same range of $f^{\prime}$ is depicted in Fig.~\ref{bif}(b).
The bifurcation scenario of each pendulum in a ring of $N=3$ coupled pendula 
is almost identical to that  of a single uncoupled pendulum (Fig.~\ref{bif}(a)) 
and the network (ring of $N=3$ coupled pendula) as a whole
exhibits a positive largest Lyapunov exponent as shown in Fig.~\ref{bif}(b).

It is to be noted that the network of diffusively coupled $(N=3)$ 
pendula is already in a synchronized state for the chosen value of coupling strength, $C=0.5$.
Consequently, following a reasoning similar to what reported in Ref.~\cite{sbjk2002}
for a system of two coupled chaotic oscillators, one gets that the synchronization
manifold has only a single positive Lyapunov
exponent for appropriate values of $f^{\prime}$.
The synchronization manifold in this case is almost similar to the phase space of a
single system (Fig.~\ref{bif}(a)) as is evident from the bifurcation diagram  (Fig.~\ref{bif}(b)). Hence 
the network as a whole exhibits a single positive Lyapunov exponent for $C=0.5$.
More details on synchronization manifold and its relation to the transition of 
Lyapunov exponents of diffusively coupled systems 
can be found in Ref.~\cite{sbjk2002}.

\begin{figure*}
\centering
\includegraphics[width=2.0\columnwidth]{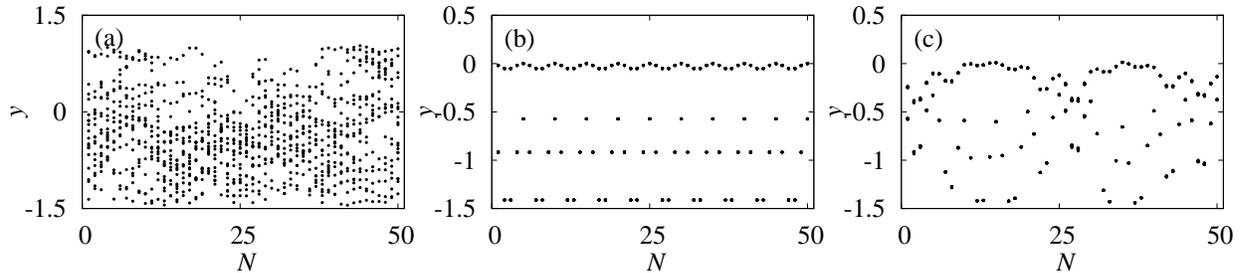}
\caption{\label{poin_nd} Poincar\'e points corresponding to the network of pendula 
in a ring configuration with $N = 50$ for the coupling strength $C=0.5$
in the absence of coupling delay.
(a) Chaotically oscillating pendula for $f^\prime=1.5$ when no disorder is present, 
(b) Periodically oscillating pendula for $20\%$ disorders with fixed $f^\prime=f^\prime_d=0.5$ and
(c) Periodically oscillating pendula for $20\%$ disorders with 
random $f^\prime_d \in(0.1,0.5)$.}
\end{figure*}
\begin{figure}
\centering
\includegraphics[width=1.0\columnwidth]{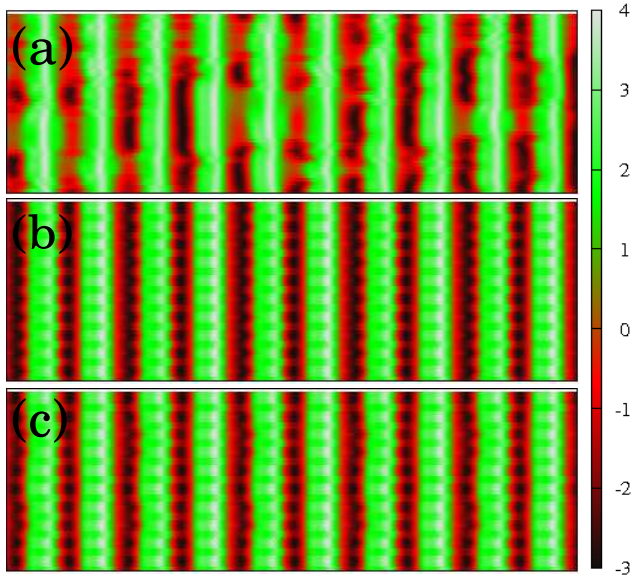}
\caption{\label{sp_nd}(Color online) Spatiotemporal representation of Fig.~\ref{poin_nd}. Here
the horizontal axis corresponds to time $t$ and the vertical axis to the oscillator index $N$.}
\end{figure}
Poincar\'e (surface of section) points corresponding to the network of $N = 50$  pendula 
in a ring configuration, (Eq.~(\ref{nodelay_eqn1})), for the coupling strength $C=0.5$ 
is depicted in Fig.~\ref{poin_nd}.  The entire network of pendula exhibits
coherent chaotic oscillations  in the absence of 
any periodically oscillating disorder 
as shown in  Fig.~\ref{poin_nd}(a) for the ac torque $f^{\prime}=1.5$. 
The spatiotemporal representation of Fig.~\ref{poin_nd}(a) is illustrated in Fig.~\ref{sp_nd}(a),
where the horizontal axis corresponds to time $t$ and the vertical axis to the oscillator index $N$,
which is plotted for ten drive cycles after leaving out sufficient transients (one thousand drive cycles).
It is to be noted that the network of $N=50$ coupled pendula does not exhibit synchronous 
chaotic oscillations as is evident from Fig.~\ref{poin_nd}(a). Otherwise it would show
identical color for all the oscillators as a function of time.
The colors code the angular velocities of the pendula; dark red (dark gray) indicates negative velocities
and green (light gray) positive velocities. Narrow bands of red and green colors represent sudden rapid
motion of the pendula in the array. The spatiotemporal plot (Fig.~\ref{sp_nd}(a)) shows that the evolution 
is not only nonperiodic but is in fact chaotic without any repetitive patterns or regular structures.

Next, impurities (disorders) with periodic oscillations are symmetrically distributed
in the array to investigate the effect of disorder as 
in the earlier studies~\cite{braiman1995,brandt2006,braimanditto1995,qi2003}.  
It is to be noted that an asymmetric distribution of disorder does not tame the 
array thereby fostering synchronous evolution and spatiotemporal regularity as 
discussed in~\cite{braiman1995}. The density of the disorder is increased
from $1\%$ along with the coupling strength $C$. We find that for $C=0.5$ 
the entire array gets locked to a synchronous periodic evolution (Fig.~\ref{poin_nd}(b))
for $20\%$ impurities with their corresponding $f^{\prime}_d=0.5$ (so that the
impurities oscillate periodically), leading to spatiotemporal
regularity (Fig.~\ref{sp_nd}(b)). Hereafter, we denote the ac torque corresponding to
chaotic states as $f^\prime$ and that corresponding to (disordered) periodic states as $f^\prime_d$. 
The spatiotemporal plot indicates repetitive 
patterns for every two drive cycles confirming the periodic evolution of the array of
pendula. To be precise, for $20\%$ disorder in the network of $N=50$ coupled pendula, 
ten disorders are placed at every fifth site in the network.
The disorder-induced spatial synchronized states reported here are exactly
for the same value of $C$ and the density of the  disorders but for different 
sizes of the array as reported in Refs.~\cite{braiman1995,brandt2006,braimanditto1995,qi2003}.  Further, 
we find that even for a random distribution of $f^{\prime}_d$ of the disorders 
taming can be achieved in a wide range of ac torque.
A periodically oscillating array of pendula for 
$20\%$ of disorder is obtained for a random distribution of $f^{\prime}_d\in(0.1,0.5)$
as shown in Fig.~\ref{poin_nd}(c) along with their spatiotemporal
representation in Fig.~\ref{sp_nd}(c).

\begin{figure*}
\centering
\includegraphics[width=2.0\columnwidth]{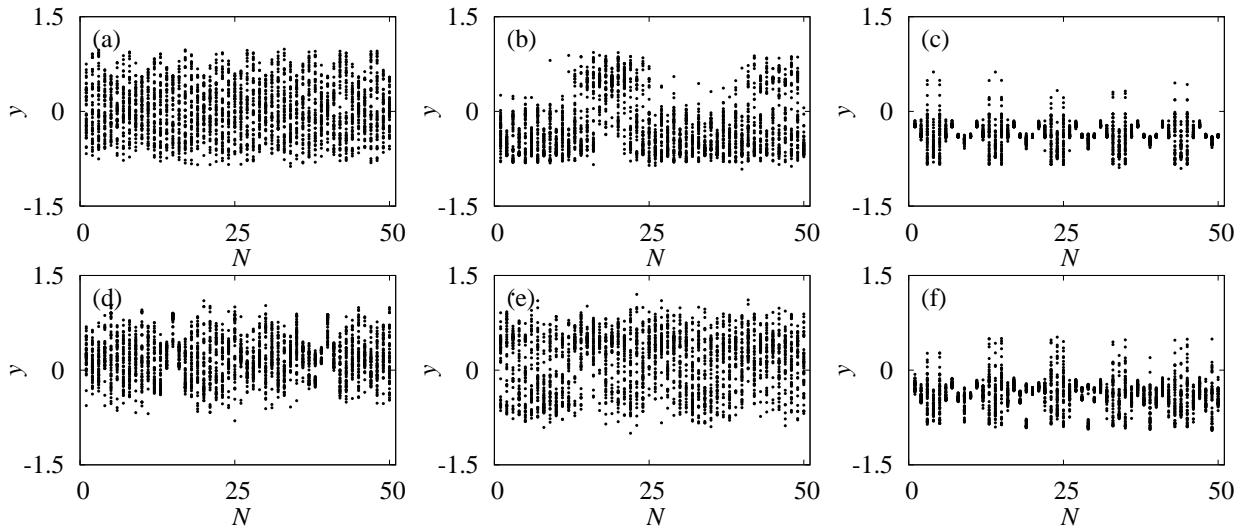}
\caption{\label{poin_t1} Same as in Fig.~\ref{poin_nd}  but now in the presence of
 coupling delay $\tau=1.5$ and for different densities of disorders. First row is with
fixed value of the disorders $f^\prime_d=0.5$ and the second row with
random values of $f^\prime_d\in(0.2,0.9)$. (a,d) $20\%$, (b,e) $50\%$ and
(c,f) $70\%$ of disorders.}
\end{figure*}
\begin{figure*}
\centering
\includegraphics[width=2.0\columnwidth]{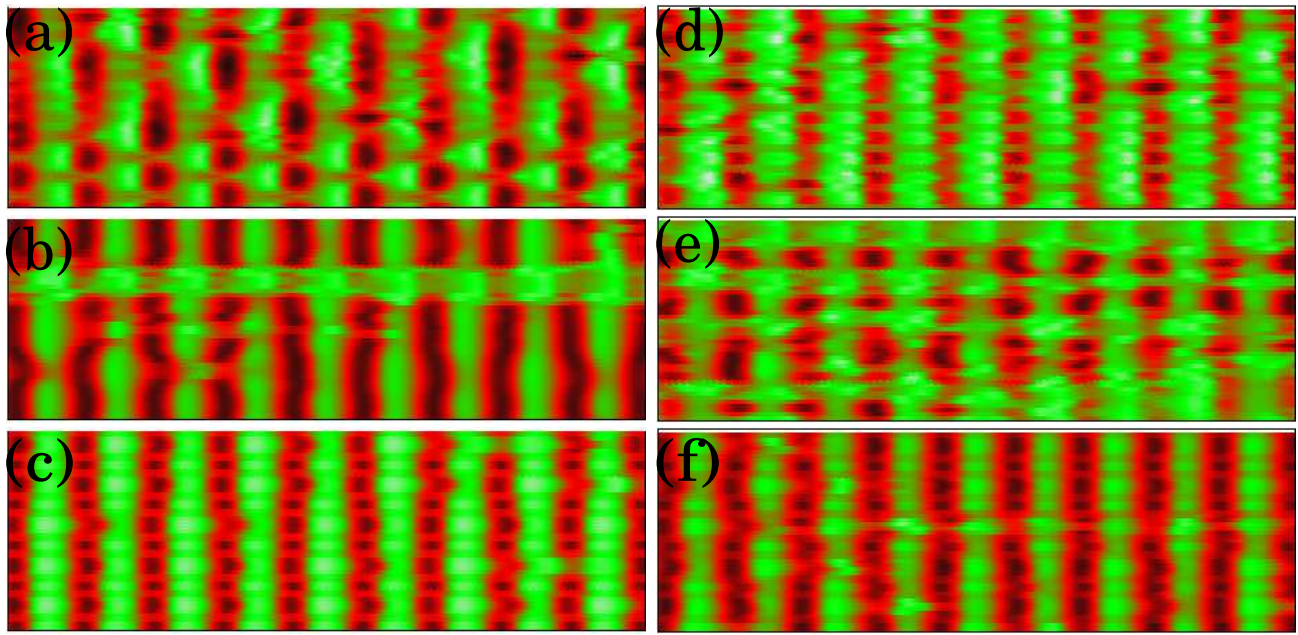}
\caption{\label{sp_t1}(Color online) Spatiotemporal representation of Fig.~\ref{poin_t1}.
Color bar is the same as in Fig.~\ref{sp_nd}.}
\end{figure*}
In the following, we will demonstrate that the introduction of coupling 
delay can sustain and enhance coherent chaotic
oscillations in the linear array with periodic boundary conditions
with the density of disorder as large as $50\%$ for the same
parameter values.  For $50\%$ disorder in the network of $N=50$ coupled pendula, 
$25$ disorders are placed at every alternate sites in the network. 
Furthermore, the coupling delay increases
the robustness of the network by preserving the dynamical complexity of 
the given network for further increase in the density of disorder to as 
large as $70\%$ of the size of the network. The array attains
synchronous periodic behavior for disorder greater than $70\%$.
For $70\%$ disorders in the network of $N=50$ coupled pendula, 
after placing $25$ disorders at every alternate sites in the network, the
remaining ten disorders are placed anywhere either symmetrically or
asymmetrically. 

\section{\label{sec:level2}Linear array with coupling delay}
\subsection{Effect of Time-delay}
Now, we consider a chain of $N$ forced coupled nonlinear pendula with
periodic boundary conditions along with
the same parameter values as in Sec.~\ref{sec:nd} but with
the  introduction of coupling delay. The dynamical equations then become,
\begin{subequations}
\begin{eqnarray}
ml^2\dot{x_i} & = & y_i,\\
\dot{y_i} & = & - \gamma y_{i}-mgl~sin~x_{i}+f+f^{\prime}_i sin(\omega t)+ \nonumber\\
& &C[y_{i+1}(t-\tau)-2y_{i}(t)+y_{i-1}(t-\tau)],
\end{eqnarray}
\label{eqn1}
\end{subequations}
where $i = 1,2,\cdots,N$ and $\tau$ is the coupling delay. 
Now, the first pendulum (see Fig.~\ref{fig1}) is 
connected with the second and with the $N^{th}$ pendulum with a delay $\tau$, so that each pendulum 
gets two delayed inputs from its nearest pendula. Similar delayed couplings
are effective for the other pendula in the array.
For $C = 0.0$, the pendula are uncoupled and evolve according to their own dynamics as before. 
As the coupling delay will change the bifurcation scenario of the coupled pendula
as a function of the ac torque, we have to look at the bifurcation diagrams to fix
the values of the strength of the ac torque
$f^\prime$ in the periodic and chaotic regimes. The bifurcation scenario 
 of a single pendulum in a ring of
$N=3$ delay coupled pendula  and the largest Lyapunov exponent of the
entire network for the value of the coupling delay $\tau=1.5$ and
for $C=0.5$ is depicted in Fig.~\ref{bif}(c). 
This network exhibits only a single positive Lyapunov
exponent for the chosen value of delay $\tau=1.5$,
as the network of diffusively coupled subsystems are synchronized
to a common synchronization manifold as discussed in Sec.~\ref{sec:nd}.
Figure~\ref{bif}(d) shows
the bifurcation diagram and its corresponding largest Lyapunov exponent
for $\tau=2.0$ and $C=0.6$. 

The bifurcation diagram (Fig.~\ref{bif}(c)) for $\tau=1.5$ and $C=0.5$
shares some common regimes of chaotic behavior in $f^\prime$
with its corresponding undelayed case (Fig.~\ref{bif}(b)). Therefore,
we fix $f^\prime=1.5$ for the chaotic pendula and $f^\prime=f^\prime_d=0.5$ for 
disorder characterized by periodic behavior as in Sec.~\ref{sec:nd}.
The Poincar\'e points 
as a function of the oscillator index (N) after leaving out a sufficiently large number of
($1000$ drive cycles) transients  in the presence of the coupling
delay, $\tau=1.5$, and for $C=0.5$ are shown in Fig.~\ref{poin_t1} for different values of
density of disorder. The first row is plotted for disorders with fixed $f^\prime_d$
and the second one for a random distribution of $f^\prime_d\in(0.2,0.9)$. The corresponding
spatiotemporal plot is depicted in Fig.~\ref{sp_t1} for 10 drive cycles. Disorder
 of size $20\%$ are uniformly distributed in the array of $N=50$ pendula,
as in Fig.~\ref{poin_nd}(b) (where the array acquired synchronous periodic oscillations)
with fixed and randomly distributed $f^\prime_d$ in the periodic regime. The
evolution of the array in  this case is illustrated in 
Figs.~\ref{poin_t1}(a) and~\ref{poin_t1}(d), respectively. The array
self-organizes to exhibit complex spatiotemporal patterns 
(Figs.~\ref{sp_t1}(a) and~\ref{sp_t1}(d)) 
without any repetitive patterns thereby exhibiting delay-induced 
phase-coherent chaotic oscillations (see Sec. \ref{phasecoh} below for confirmation). 
It is to be noted that the network of  $N=50$ delay coupled
pendula does not exhibit synchronous oscillations as confirmed by 
the Figs.~\ref{sp_t1}(a) and~\ref{sp_t1}(d).
We have increased the density of disorder
upto $50\%$ for the same values of the parameters and the scenario is depicted
in Figs.~\ref{poin_t1}(b) and~\ref{poin_t1}(e) for fixed and random 
distributions of $f^\prime_d$, respectively,
along with the spatiotemporal representation in Figs.~\ref{sp_t1}(b) and~\ref{sp_t1}(e).
These figures show that the array of pendula originally with $50\%$ of periodic disorder
evolves to acquire collective coherent chaotic oscillations in the entire array induced by the 
rather small coupling delay $\tau=1.5$ in a wide range of $f^\prime$, which 
is indeed a surprising result of delay impact. 
\begin{figure*}
\centering
\includegraphics[width=2.0\columnwidth]{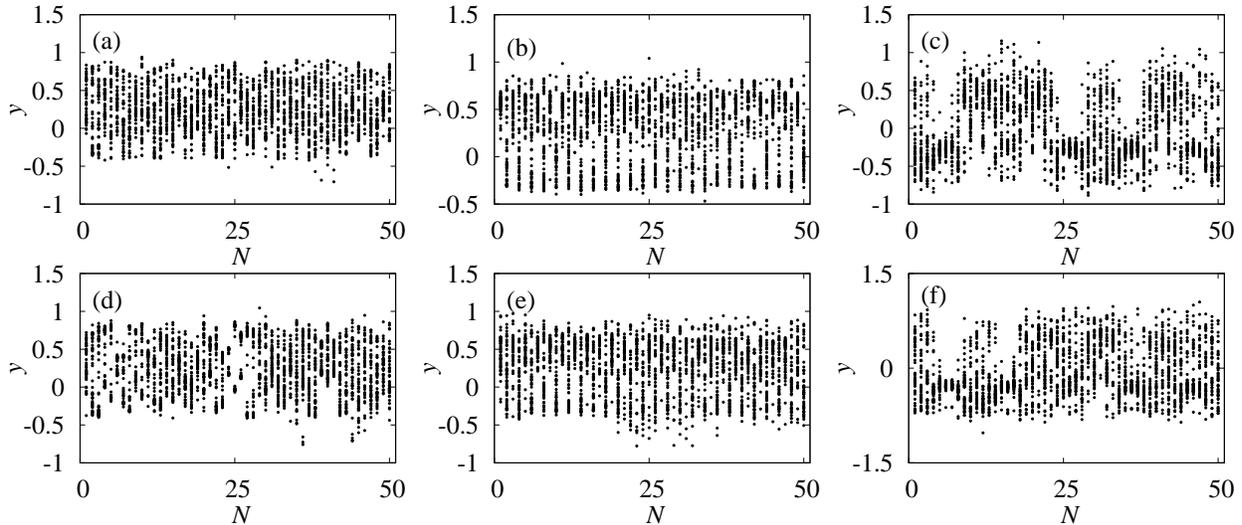}
\caption{\label{poin_t2} Poincar\'e points corresponding to the network of pendula, 
with $f^\prime=1.6$ for chaotic pendula,
in a ring configuration with $N = 50$ for the coupling strength $C=0.6$
with  coupling delay $\tau=2.0$ and for different values of density of disorders.
First row is with
fixed value of the disorders $f^\prime_d=1.1$ and the second row with
random values of $f^\prime_d\in(0.5,1.2)$. (a,d) $20\%$, (b,e) $50\%$ and
(c,f) $70\%$ of disorders.}
\end{figure*}
\begin{figure*}
\centering
\includegraphics[width=2.0\columnwidth]{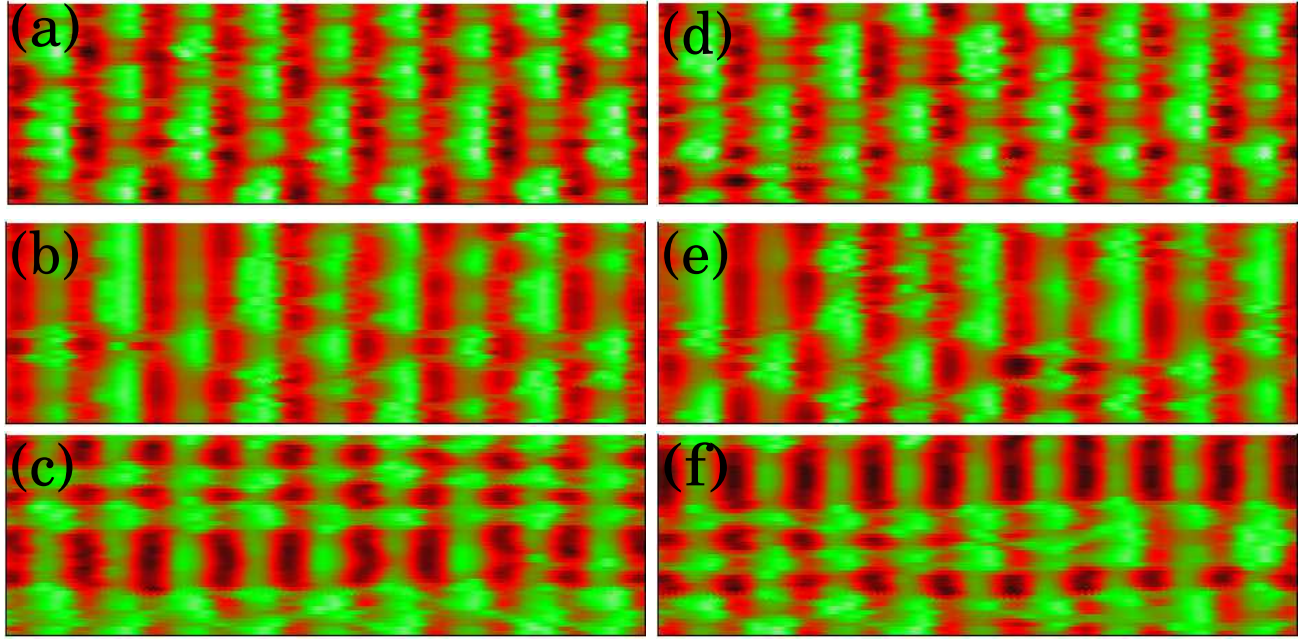}
\caption{\label{sp_t2} (Color online) Spatiotemporal representation of Fig.~\ref{poin_t2}.
Color bar is the same as in Fig.~\ref{sp_nd}.}
\end{figure*}
\subsection{Delay enhanced phase coherence}
\label{phasecoh}
\begin{figure}
\centering
\includegraphics[width=1.0\columnwidth]{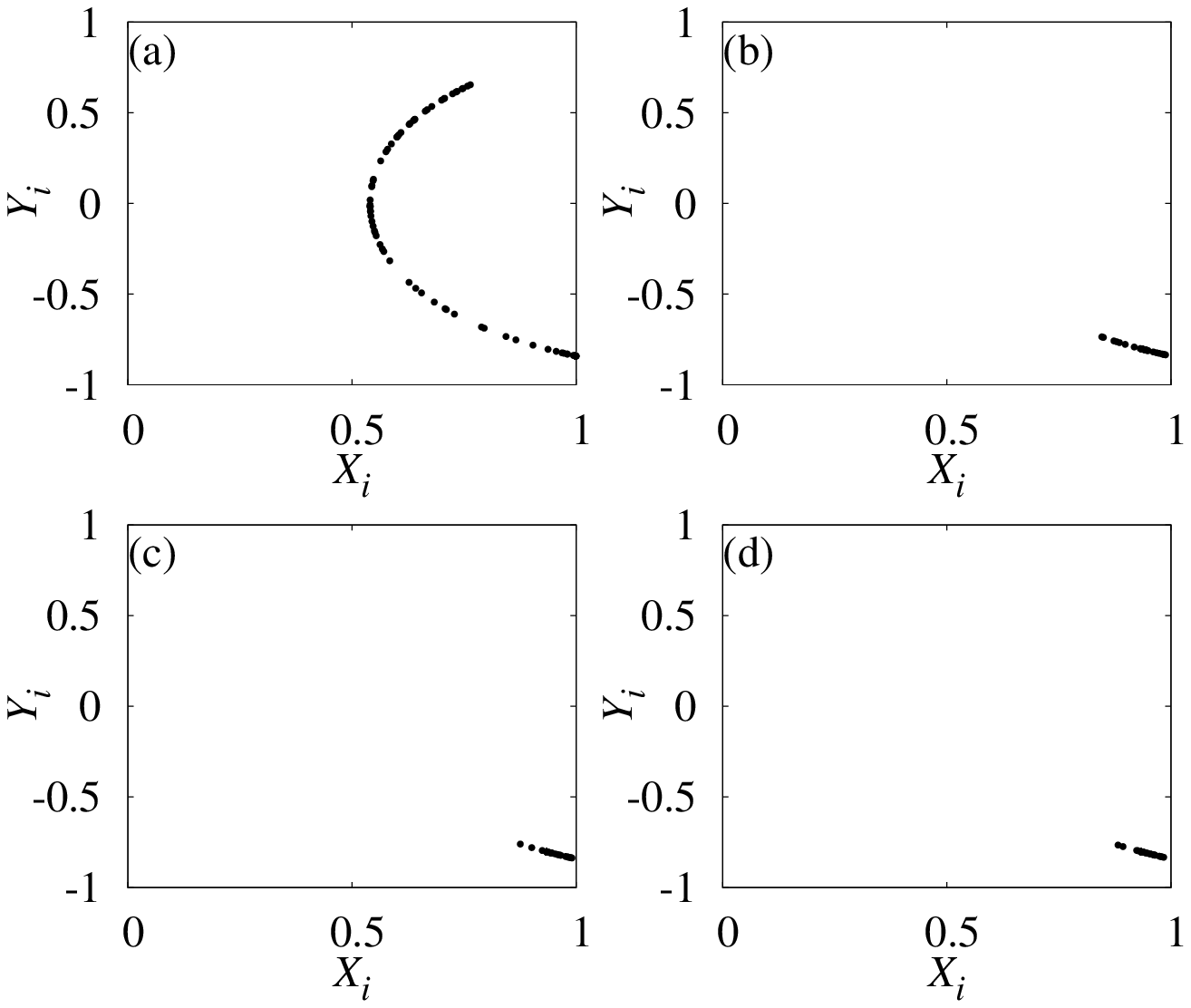}
\caption{Snapshots of the phase portraits on the $(X_i,Y_i)$ plane of the ring network, 
with $X_i=\cos \theta_i$ and $Y_i=\sin \theta_i$. Here (a) corresponds 
to $C=0.05$ and (b)-(d) correspond to Figs.~\ref{poin_t1}(d)-(f) with $C=0.5$ and the strength
of the impurities being $20\%, 50\%$ and $70\%$, respectively.}
\label{linear_cup_ps}
\end{figure}

Controlling of oscillator coherence by delayed feedback has been observed both theoretically
and experimentally in Refs.~\cite{dgmr2003,sbea2004}. In our investigation, we find that in addtion to the
enrichment of the periodic disorder to (chaotic) higher order oscillations, delay
coupling also increases the coherence of the collective chaotic oscillations of the whole network.
For a better understanding and confirmation of the delay enhanced phase-coherent 
oscillations of the entire network, we investigate both qualitatively and quantitatively
the coherence property of the network macroscopically.
For each of the pendulum in system (\ref{eqn1}), one can define the phase as
\begin{eqnarray}
\theta_i=\tan^{-1}(y_i/x_i).
\label{phase1}
\end{eqnarray}
Here $\theta_i$'s represent the phases of the individual pendula in the system. 
In order to visualize the effect of coupling delay on phase coherence of the system, 
we plot the snapshot of the phases of the pendula in Fig.~\ref{linear_cup_ps}. 
From Eq.~(\ref{phase1}), one can write
\begin{subequations}
\begin{eqnarray}
X_i&=&\cos\theta_i=\frac{x_i}{\sqrt{(x_i^2+y_i^2)}},\\
Y_i&=&\sin\theta_i=\frac{y_i}{\sqrt{(x_i^2+y_i^2)}}.
\end{eqnarray}
\label{phase2}
\end{subequations}
The Kuramoto order parameter $r$ which quantifies the strength of phase coherence 
is given by $re^{i\psi}$=$\frac{1}{N}\sum_{j=1}^N e^{i\theta_j}$. 
When $r=0$ phase coherence is absent in the system and when $r \approx 1$ 
there is complete phase coherence in the system. Thus $r$ essentially 
quantifies the strength of phase coherence. 
To be more quantitative one can use the
time averaged order parameter $R=\frac{1}{T}\int^{T}_{0} rdt$ so that 
its low value (near to zero) corresponds to phase incoherence while a value near to 
unity corresponds to phase coherence. Throughout the manuscript, 
we have estimated $R$ for an average over $1200$ time units.

\begin{figure}
\centering
\includegraphics[width=1.0\columnwidth]{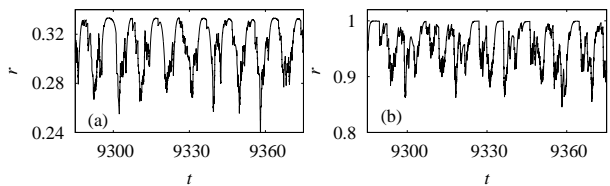}
\caption{Time evolution of the Kuramoto order parameter $r$ for
(a) $C=0.05$ (Fig.~\ref{linear_cup_ps}(a)) and (b) $C=0.5$ with $20\%$ impurity
(Fig.~\ref{linear_cup_ps}(b)).}
\label{linear_pend_ts_op}
\end{figure}

Using Eqs.~(\ref{phase2}) (where the $x_i$'s are wrapped to be between $0$ and $2\pi$) 
we have plotted the distribution of phases associated with Eq.~(\ref{eqn1}) in the $(X_i,Y_i)$ plane 
in Fig.~\ref{linear_cup_ps}. We present the results for two specific values of the 
coupling strength $C$ for the same value of delay parameter $\tau=1.5$, that is for a low value 
of coupling $C=0.05$ with $20\%$ impurity (Fig.~\ref{linear_cup_ps}(a)) and for
$C=0.5$ with $20\%, 50\%$ and $70\%$ impurities in Figs.~\ref{linear_cup_ps}(b), (c) and 
(d), respectively. The phases of the pendula are distributed apart on
the unit circle for $C=0.05$, as illustrated in Fig.~\ref{linear_cup_ps}(a), indicating a poor or a very low
coherence of the pendula, which is  also confirmed by the low value of the time average 
of the Kuramoto order parameter $R=0.316$. 
The time evolution of the corresponding order parameter $r$ itself is
depicted in Fig.~\ref{linear_pend_ts_op}(a).
The phases of the pendula in the entire network 
corresponding to Figs.~\ref{poin_t1}(d)-(f), that is for $C=0.5$ with $20\%, 50\%$ and $70\%$ impurities,
are depicted in Figs.~\ref{linear_cup_ps}(d)-(f), respectively.
The phases are now confined to a much smaller region on the unit circle 
for $C=0.5$ in the presence of the delay coupling confirming the delay enhanced 
phase-coherent oscillations of the entire network. 
This is indeed
confirmed by much higher values of the time averaged order parameter $R=0.964, 0.973$, and $0.982$, 
for Figs.~\ref{linear_cup_ps}(b)-(d), respectively. 
Also, the time evolution of the order parameter corresponding to Fig.~\ref{linear_cup_ps}(b)
is shown in Fig.~\ref{linear_pend_ts_op}(b).
Thus, we have confirmed the existence of delay enhanced phase-coherent oscillations 
in the entire network of delay coupled pendula for appropriate coupling strength $C$.

\subsection{Possible Mechanism}
\label{mech}
Two simple mechanism were suggested for taming of chaos by disorder and fostering of
periodic patterns in the array without delay in Ref.~\cite{braiman1995}. 
Indeed, we find the manifestations of both
these mechanism in the delay coupled networks as well under appropriate conditions.  
It is essential to understand
the first of these two mechanism to understand the mechanism behind the delay
induced coherent chaotic oscillations. The first mechanism depends on the 
topological features of the attractors. The periodic disorder needed to stabilize a
chaotic array depends on both the distance and the direction in the
parameter space to the nearest periodic attractor, which is controlled by the
magnitude and distribution of the disorder~\cite{braiman1995}. For this mechanism 
to work it is not essential to introduce disorder since uncoupled chaotic oscillators
can become periodic when coupled. This phenomenon is explicitly observed from the
bifurcation diagrams shown in Fig.~\ref{bif}. The chaotic regimes in the
range of $f^\prime\in(0.5,0.97)$ in Fig.~\ref{bif}(a) for the uncoupled system
becomes periodic for the same parameters when a coupling delay is introduced (see Fig.~\ref{bif}(c)
and ~\ref{bif}(d)). The second mechanism
deals with the locking of the chaotic pendula by the periodic ones to the 
external ac drive~\cite{braiman1995}, which is observed for disorder greater than
$70\%$ in our case.

\begin{figure}
\centering
\includegraphics[width=0.7\columnwidth]{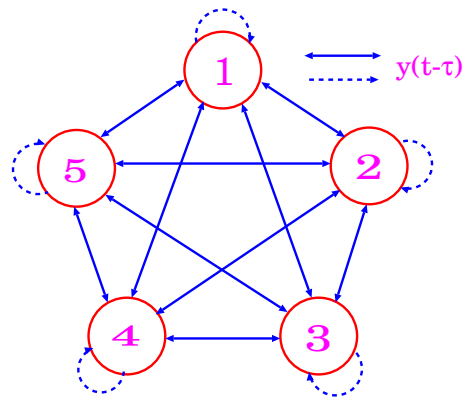}
\caption{\label{fig5} (Colour online) Schematic diagram of a network of $5$ 
oscillators with all to all (global) coupling. In this figure each oscillator 
gets $4$ delayed input from the remaining oscillators in the network, 
and also a delayed feedback from itself when $j = i$.}
\end{figure}
In this paper, we are interested in delay-induced coherent chaotic oscillations
in the network, the mechanism of them is a simple extension of
the first, as will be explained in the following.
The presence of delay in the coupling extends the phase space
dimension as a time-delay system is essentially an infinite-dimensional
system~\cite{dvsbook}. Therefore, the dimension and the phase space  
(characterized by multiple unstable directions corresponding to multiple 
positive Lyapunov exponents of the delay-coupled network) of the chaotic attractors
of the delay-coupled network also increases. This in turn increases the robustness 
of the chaotic attractors against even nearby periodic orbits (disorders)
in the parameter space  and hence the presence of a large percentage of periodic disorder
(which does not extend over multi-dimensional phase space) is not capable of 
taming the chaoticity of the pendula. Furthermore, delay being 
a source of instability, by inducing
chaotic oscillations~\cite{pwrsid1993,nkgv1995,jmalb1987,kg1992,dvsbook}, the
periodic disorder acquires chaotic oscillations for suitable values
of the delay resulting in coherent chaotic oscillations of the entire array.
It is to be noted that increasing the delay alone gives rise to a rich variety of 
behavior, such as periodic, higher order oscillations, chaotic and hyperchaotic
attractors with a large number of positive Lyapunov exponents as observed in 
several bifurcation diagrams presented earlier as a function of the delay even in scalar 
time-delay systems~\cite{dvsbook,jdf1982}.
Further, periodic orbits of very large periods are also created due to the delay which are
not present in the undelayed systems and these higher order oscillation
manifest in the array  in place of disorder when the size of disorder 
is larger than $50\%$.

\begin{figure*}
\centering
\includegraphics[width=2.2\columnwidth]{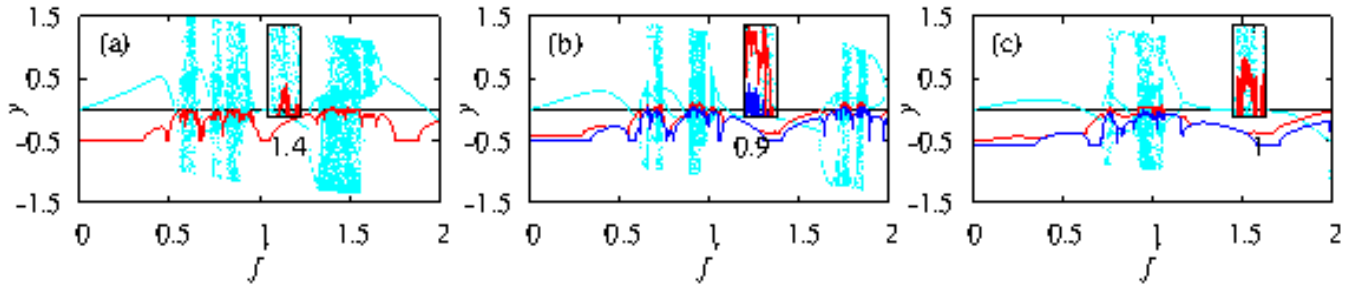}
\caption{\label{bif_gc}(Color online)  The largest Lyapunov exponents of the network of 
$N=20$ globally coupled pendula and the bifurcation diagram of a single pendulum in the network 
for different values of the coupling strength $C$ and the coupling delay $\tau$. The inset
shows that the network exhibits chaotic oscillations for the specific values of
$f^\prime$ chosen in the text.
(a) $C=0.2$ and $\tau=0.0$,
(b) $C=0.2$ and $\tau=1.5$, and
(c) $C=0.3$ and $\tau=2.0$. Red and dark blue (dark grey) lines correspond to the largest 
Lyapunov exponents and light blue (light grey) dots correspond to the bifurcation diagram.}
\end{figure*}
\subsection{Effect of Increased Disorder and Coupling Delay}
We have also increased the amount of disorder to more than half the size of
the network to investigate the effect of delay coupling. Indeed this scenario may also
be considered as the one in which chaotic impurities coexist in a sea of a periodically oscillating network and
one may expect the suppression of chaotic oscillations to achieve coherent periodic
oscillations so as to enhance spatiotemporal order of the network.  Nevertheless, the presence
of delay in the coupling prohibits suppression of any  chaotic pendula
 upto $70\%$ of disorder and induces
chaoticity in the periodic impurities adjacent to a chaotic pendulum, while the
impurities away from it acquire higher order oscillations.
Furthermore, it is to be noted that for a density of disorder larger than $50\%$,
the distribution of disorder becomes nonuniform (asymmetric). 
For instance, we have distributed $70\%$ of disorders, while retaining 
chaoticity only in the remaining $30\%$ pendula for the same values of
$\tau$ and $C$. The dynamical organization of the array with $70\%$ of disorder
is shown in Figs.~\ref{poin_t1}(c) and~\ref{poin_t1}(f), which again
depicts the delay-induced chaoticity in the
periodic pendula adjacent to the chaotic pendulum, while the other periodic
pendula away from the chaotic pendulum acquire higher order oscillations. 
The corresponding self-organized complex spatiotemporal behavior is 
shown in Figs.~\ref{sp_t1}(c) and~\ref{sp_t1}(f).
We have also confirmed the higher order oscillations of the pendula from their
corresponding phase space plots.
We can conclude that the complexity of the network as a whole is
increased in the presence of delay coupling even if the impurities exceed half
the size of the network thereby confirming the robustness of
the network against large disorder-induced by the coupling delay.

Next, the value of the coupling delay is further increased to examine whether
the delay enhances the coherent chaotic oscillations and increases the robustness 
of the array against more than $70\%$  disorder. We find that increase in
the coupling delay also leads to the same results for appropriate value
of the coupling strength and the network of pendula attains synchronous periodic 
oscillations leading to spatiotemporal order for disorder of size more than $70\%$.
To be specific, we fix the coupling delay as $\tau=2.0$ and $C=0.6$. 
For impurities of periodic type the ac torque is fixed as $f^\prime_d=1.1$ and for
chaotic pendula it is chosen as $f^\prime=1.6$ using the bifurcation diagram shown in
Fig.~\ref{bif}(d) (as seen in the inset). The first row in Fig.~\ref{poin_t2} is plotted for 
disorders with fixed $f^\prime_d$
and the second row for random distribution of $f^\prime_d\in(0.2,0.9)$.
Delay-induced coherent chaotic oscillations
of the whole network in the presence of $20\%$ disorder
are shown in Figs.~\ref{poin_t2}(a) and~\ref{poin_t2}(d). The
corresponding spatiotemporal representation is illustrated in Figs.~\ref{sp_t2}(a) and~\ref{sp_t2}(d).
The network of $N=50$ coupled pendula oscillates chaotically even
in the presence of $50\%$ disorder as depicted in Figs.~\ref{poin_t2}(b) and~\ref{poin_t2}(e)
along with their complex spatiotemporal patterns in Figs.~\ref{sp_t2}(b) and~\ref{sp_t2}(e), respectively.
Further increase in the density of disorder to $70\%$  continues to result in chaotic
oscillations of disorders adjacent to the chaotic pendula and higher order oscillations
in disorders further away from it, as shown in Figs.~\ref{poin_t2}(c) and~\ref{poin_t2}(f).
The corresponding dynamical organization of the network of pendula with $70\%$ disorder
to self-organized complex spatiotemporal structures is illustrated in
Figs.~\ref{sp_t2}(c) and~\ref{sp_t2}(f).

\subsection{Summmary}
Thus we have shown that the infected sites are healed or in other words
the disorders in the ring network acquires coherent chaotic oscillating behavior induced
by time-delay in the coupling thereby enhancing the spatiotemporal complexity for a uniform (symmetric)
distribution of the impurities as large as $50\%$ of the array. 
Note that in the absence of delay in the coupling, the whole network will
become infected (ordered) even for $20\%$ of disorder. Further for the density 
of disorder larger than $50\%$,
the distribution of disorder becomes nonuniform (asymmetric). In this case, the
impurities adjacent to the chaotic element acquires chaoticity, while the impurities 
located away from the chaotic ones acquire higher order oscillations 
resulting in enhanced complexity of the network. It is also to
be appreciated that the delay in the coupling not only enhances the coherent chaotic 
oscillations, but also increases the robustness of the
network against any infection (disorder) of even more than half the size of the network.

In the next section, we will extend our investigation to a network of globally coupled
pendula and show that we essentially obtain similar results. In particular, coupling
delay can enhance the dynamical complexity of disordered pendula leading to 
delay-induced coherent chaotic oscillations upto $50\%$ of symmetric disorder. 
For asymmetric disorder of size greater than $50\%$, the coupling delay can 
induce chaotic oscillations in disordered pendula adjacent to chaotic pendula
and those away from it will acquire higher order oscillations upto $65\%$
disorder resulting in the enhanced complex behavior of the existing network.
%
\section{\label{sec:level3}Global delayed coupling}
Most natural systems involve complicated coupling between them and that the individual 
oscillators are not only coupled with their nearest 
neighbors but also with all other elements in the network. 
Such a global coupling plays an important role in a large number of dynamical systems 
ranging from the physical \cite{alekseev1998}, chemical \cite{middya1994} and
biological \cite{winfree1980} to social and economic \cite{gonzalez2006,matassini2001} 
networks and electronic systems \cite{visarath2003}. 
Global coupling is also being studied in reaction-diffusion systems, for example as a
reaction-diffusion with global coupling (RDGC) model, to understand the mechanism 
behind the electromechanic dynamics of the heart and generation of successive 
ectopic beats \cite{alvarez2009} 
and also to understand the mechanism behind the oscillatory regime in the Nash-Panfilov model \cite{nash2004}.
In addition, delayed global coupling has been shown to induce in-phase synchronization 
in an array of semiconductor lasers \cite{kozyreff2000}. It has been demonstrated 
that global coupling is more efficient than local coupling to achieve non-stationary 
and stationary in-phase operations with and without delay, 
respectively, in Ref.\cite{li1993,gareia1999}. 

In this section, we will investigate the effect of delay coupling in the
presence of disorder in a globally coupled network of pendula, 
where every pendulum is connected to all the other ($N-1$) pendula in the 
network with a delay $\tau$ and it gets a self-delayed feedback only 
when $j = i$. To explain the coupling configuration, 
a schematic diagram of a network of $5$ oscillators is shown in 
Fig.~\ref{fig5} (the doted lines show the 
self-delayed feedback only when $j=i$). The model is represented as
\begin{subequations}
\begin{eqnarray}
ml^2\dot{x_i} & = & y_i,\\
\dot{y_i} & = & - \gamma y_{i}-mgl~sin~x_{i}+f+f^{\prime}_i sin(\omega t)+\nonumber \\
& &\frac{C}{N} \sum_{j=1}^N [y_{j}(t-\tau) - y_{i}(t)],
\end{eqnarray}
\label{eqn2}
\end{subequations}
where $i = 1,2,\cdots,N$. All the parameters have been chosen to be
the same as in the previous section. 
We restrict ourselves to $N = 20$ oscillators for computational
convenience; however similar results have also been obtained for larger number 
of oscillators for appropriate coupling delay and  coupling strength.

Further, we wish to add that in order to fix the system parameters pertaining to
chaotic and periodic regimes, unlike the case of linear coupling (Sec.~\ref{sec:nd} and 
\ref{sec:level2}), it is not meaningful to consider the bifurcation scenario with
low numbers of pendula, like $N=3$ or $4$, in the case of global coupling as the 
bifurcation diagrams will change appreciably when the value of $N$ changes. So in our following 
study of the bifurcation scenario and the Lyapunov spectrum, we analyse  the full network itself
and present the first one or two largest Lyapunov exponents of the entire network and the bifurcation 
diagram of a single pendulum in the network.

\begin{figure*}
\centering
\includegraphics[width=2.0\columnwidth]{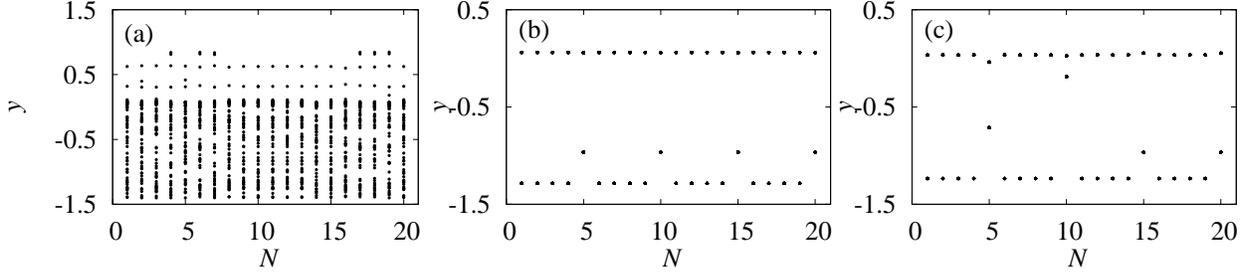}
\caption{\label{poin_gc_nd} Poincar\'e points of globally coupled network of  $N=20$ pendula
for the coupling strength $C=0.2$ in the absence of coupling delay $\tau=0.0$.
(a) Chaotically oscillating pendula for $f^\prime=1.4$,
(b) Periodically oscillating pendula for $20\%$ of disorders  with fixed $f^\prime_d=0.3$, and
(c) Periodically oscillating pendula for $20\%$ of disorders  with random $f^\prime_d\in(0,0.3)$.}
\end{figure*}
\begin{figure}
\centering
\includegraphics[width=1.0\columnwidth]{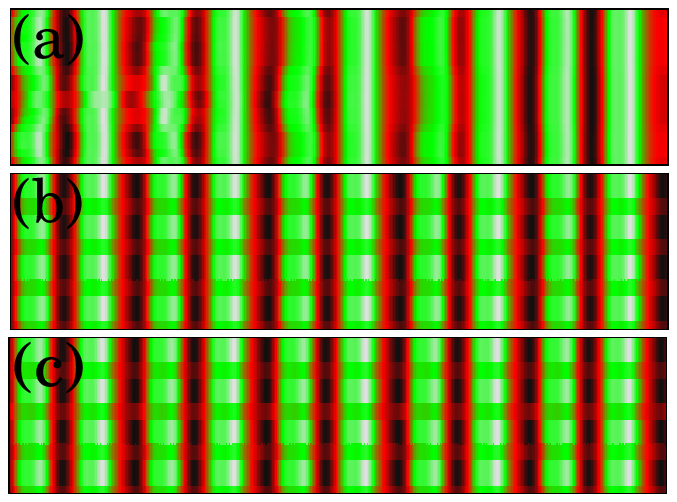}
\caption{\label{sp_gc_nd}(Color online) Spatiotemporal representation of Fig.~\ref{poin_gc_nd}.
Color bar is the same as in Fig.~\ref{sp_nd}.}
\end{figure}
\begin{figure*}
\centering
\includegraphics[width=2.0\columnwidth]{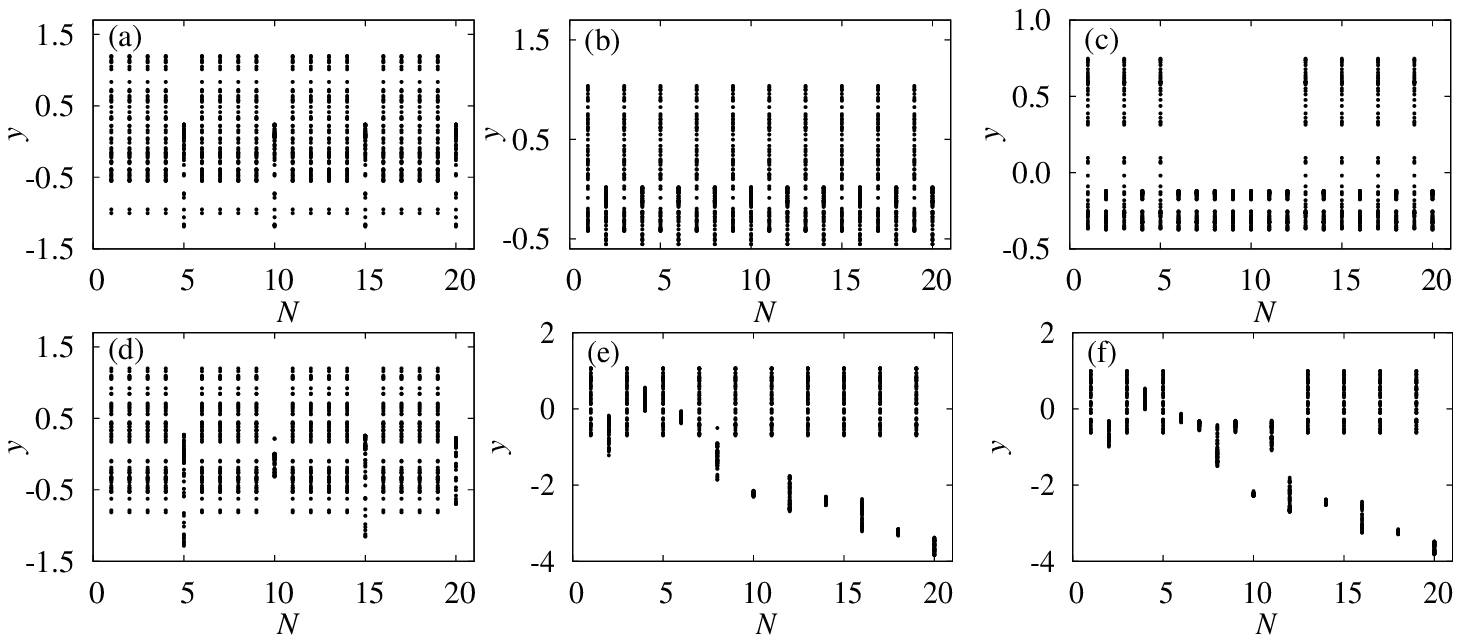}
\caption{\label{poin_gc_t1} Poincar\'e points of globally coupled network of  $N=20$ pendula,
chaotic for $f^\prime=0.92$,
for the coupling strength $C=0.2$ and the coupling delay $\tau=1.5$.
First row with
fixed value of the disorders $f^\prime_d=1.5$ and the second row with
random values of $f^\prime_d\in(1.0,1.7)$.
(a,d) $20\%$, (b,e) $50\%$ and
(c,f) $65\%$ of disorders.}
\end{figure*}
\begin{figure*}
\centering
\includegraphics[width=2.0\columnwidth]{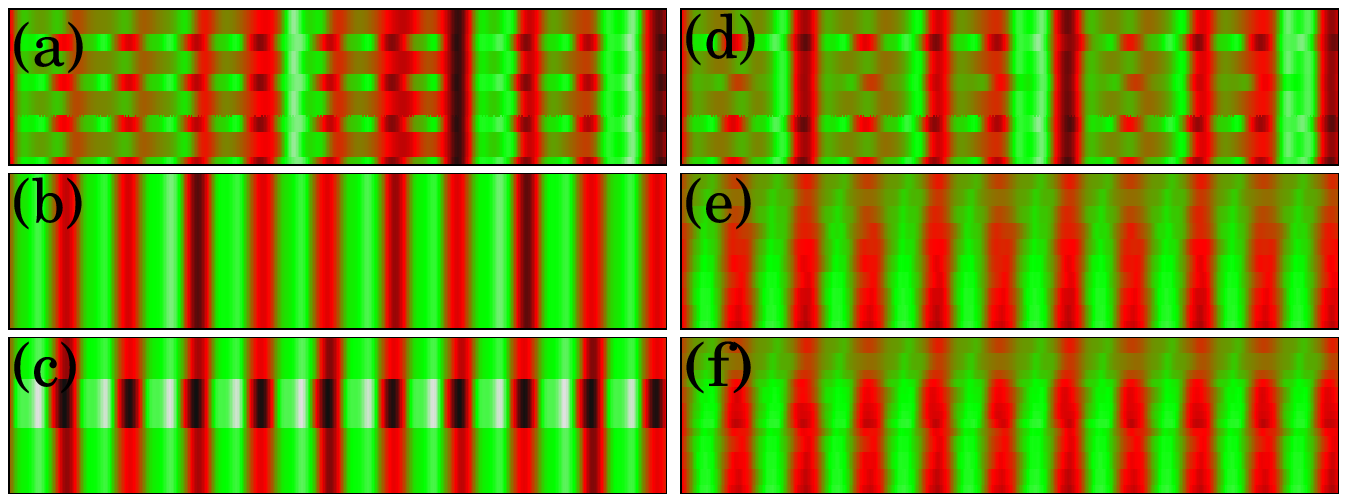}
\caption{\label{sp_gc_t1}(Color online) Spatiotemporal representation of Fig.~\ref{poin_gc_t1}.
Color bar is the same as in Fig.~\ref{sp_nd}.}
\end{figure*}
\subsection{\label{sec:gc_nd}Globally coupled pendula without delay}
We will start our investigation by plotting the bifurcation diagrams
and the Lyapunov exponents for delineating the periodic and
chaotic regimes in the case of $N=20$ globally coupled pendula. Enhancement of
spatiotemporal regularity and taming chaoticity in globally coupled
network has not yet been reported to the best of our knowledge. 
Hence the comparison of delay-enhanced
coherent chaotic oscillations leading to spatiotemporal complexity will
be meaningful only when the globally coupled chaotic network in the presence
of a few periodic disorder is tamed when there is no delay. 
Therefore, in this section, we will show that the 
globally coupled chaotic network is indeed tamed leading to 
spatiotemporal order in the absence of coupling delay.

\begin{figure*}
\centering
\includegraphics[width=2.0\columnwidth]{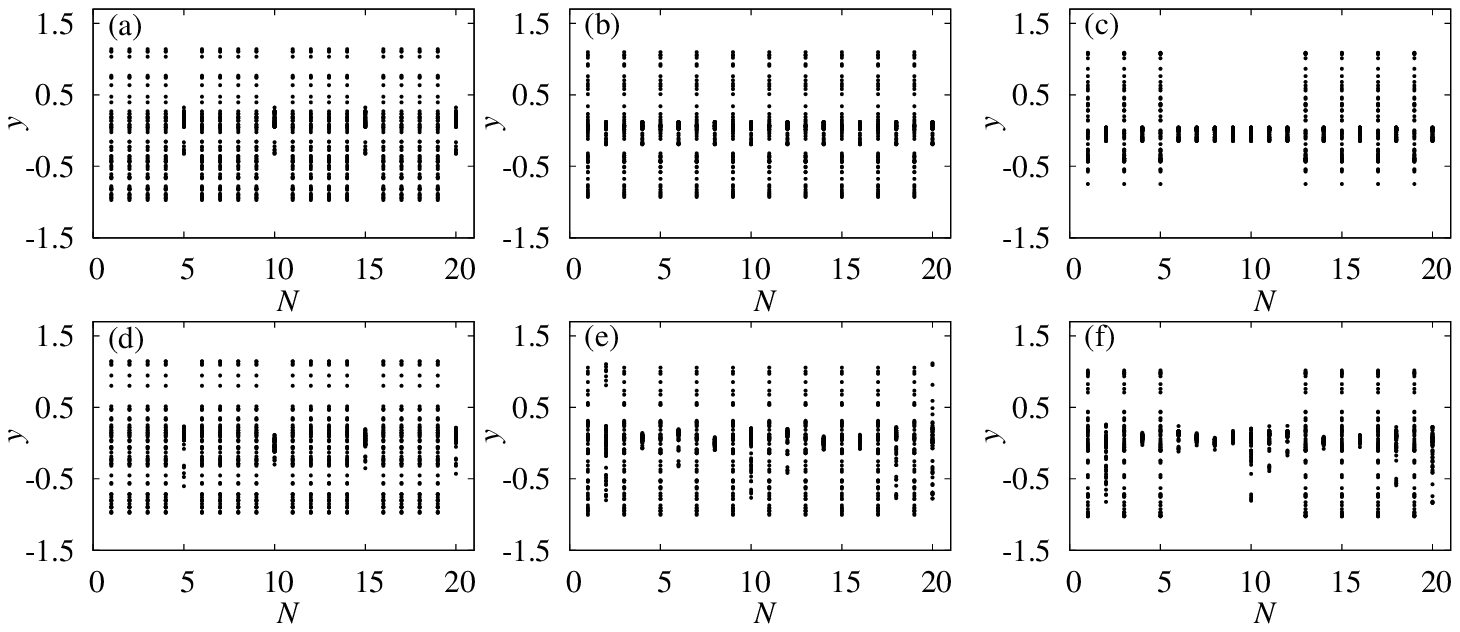}
\caption{\label{poin_gc_t2} Poincar\'e points of globally coupled network of  $N=20$ pendula,
chaotic for $f^\prime=0.96$,
for the coupling strength $C=0.3$ and the coupling delay $\tau=2.0$.
First row with
fixed value of the disorders $f^\prime_d=1.5$ and the second row with
random values of $f^\prime_d\in(1.2,2.0)$.
(a,d) $20\%$, (b,e) $50\%$ and
(c,f) $65\%$ of disorders.}
\end{figure*}
\begin{figure*}
\centering
\includegraphics[width=2.0\columnwidth]{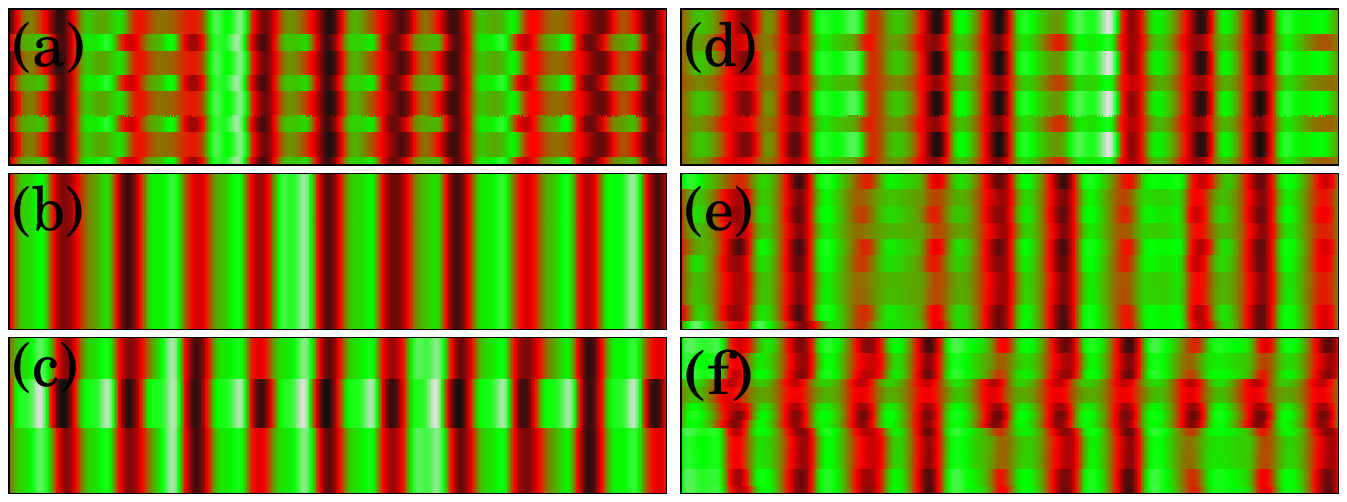}
\caption{\label{sp_gc_t2}(Color online) Spatiotemporal representation of Fig.~\ref{poin_gc_t2}.
Color bar is the same as in Fig.~\ref{sp_nd}.}
\end{figure*}
The largest Lyapunov exponent of the full network and the bifurcation diagram of a 
pendulum in the network of the globally coupled ($N=20$) case are plotted in Fig.~\ref{bif_gc}(a) for
$C=0.2$ in the range of $f^\prime\in(0,2)$ when no delay is present. We find that all the pendula in this network 
exhibit an almost similar bifurcation scenario, and that the network as a whole exhibits multiple positive 
Lyapunov exponents. However, these values are close to each other and so we present only the
largest one in Fig.~\ref{bif_gc}(a). We fix $f^\prime=1.4$ for chaotic pendula (as confirmed from
the positive Lyapunov exponent shown in the inset of Fig.~\ref{bif_gc}(a)) and $f^\prime_d=0.3$ 
for periodic disorder from the bifurcation diagram. Chaotically oscillating 
pendula in the globally coupled network without any disorder is depicted in
Fig.~\ref{poin_gc_nd}(a) for $C=0.2$ along with its
spatiotemporal representation in Fig.~\ref{sp_gc_nd}(a)
for $10$ drive cycles. As discussed in the case of diffusive coupling 
in Sec.~\ref{sec:nd}, the globally coupled network is also tamed 
exhibiting periodic oscillations for $20\%$ periodic uniform disorder with
fixed $f^\prime_d=0.3$, as illustrated in Fig.~\ref{poin_gc_nd}(b) for the
same value of $C$ in the absence of delay.  The corresponding spatiotemporal plot 
shows spatiotemporal regularity with repetitive patterns for every 
two drive cycles as depicted in Fig.~\ref{sp_gc_nd}(b). 
We have obtained similar results of taming chaoticity (Fig.~\ref{poin_gc_nd}(c)) 
from the network leading to spatiotemporal order (Fig.~\ref{sp_gc_nd}(c))
for random values of the ac torque $f^\prime_d\in(0,0.3)$. It is also to be noted
that we have also got similar results  for random values of 
$f^\prime_d\in(1.7,2.0)$.

\begin{figure}
\centering
\includegraphics[width=1.0\columnwidth]{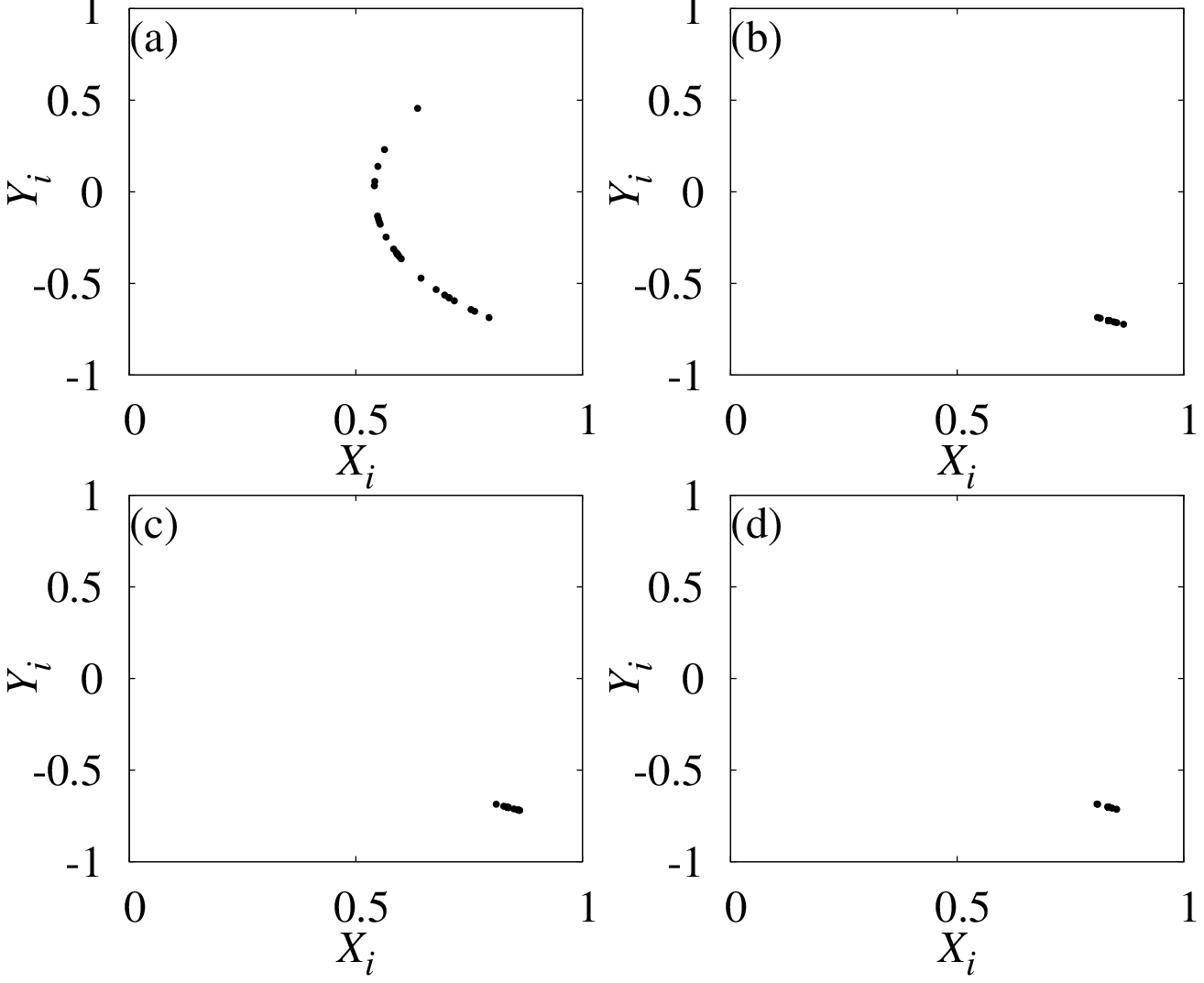}
\caption{Snapshots of the phase portraits on the $(X_i,Y_i)$ plane 
for a globally coupled network of pendula. 
Here (a) corresponds to $C=0.05$, and (b)-(d) corresponds to Figs.~\ref{poin_gc_t1}(d)-(f)
with $C=0.2$ and the strength of the impurities being $20\%, 50\%$ and $65\%$, respectively.}
\label{gloab_cup_ps}
\end{figure}

In the next section, we will demonstrate the existence of delay-induced 
coherent chaotic oscillations leading to enhanced 
spatiotemporal complexity of the network for disorder of
size as large as $65\%$.

\subsection{Globally coupled pendula with coupling delay}
The largest two Lyapunov exponents of the network of globally 
delay coupled pendula and the bifurcation diagram of a single pendulum in the network
are plotted in Fig.~\ref{bif_gc}(b)
for the same value of coupling delay and coupling strength as in the
non-delay case reported in the previous section (Sec.~\ref{sec:gc_nd}) 
for comparison. Again the system as a whole exhibits $20$ positive Lyapunov
exponents  and only the first two largest positive Lyapunov exponents differ appreciably from
the other almost identical positive Lyapunov exponents. Now, we fix $f^\prime=0.92$ 
for chaotic pendula (as confirmed from
the positive Lyapunov exponent shown in the inset of Fig.~\ref{bif_gc}(b)) and $f^\prime_d=1.5$ 
for periodic disorder from the bifurcation diagram.
Poincar\'e points representing
the delay-induced coherent chaotic oscillations of $N=20$ pendula in the network 
with $20\%$ symmetric disorder with fixed $f^\prime_d=1.5$ are illustrated in Fig.~\ref{poin_gc_t1}(a)
with its complex spatiotemporal patterns in Fig.~\ref{sp_gc_t1}(a). A random 
distribution of $f^\prime_d\in(1.0,1.7)$ corresponding to $20\%$ disorder 
also results in  delay-induced coherent chaotic oscillations (Fig.~\ref{poin_gc_t1}(d))
and enhanced spatiotemporal complexity (Fig.~\ref{sp_gc_t1}(d)). The density of
disorder is increased further upto half the size of the network
 with both fixed $f^\prime_d=1.5$
and random  distribution of $f^\prime_d\in(1.0,1.7)$ as in Figs.~\ref{poin_gc_t1}(b)
and ~\ref{poin_gc_t1}(e), respectively, which shows increased complexity of
the entire network as depicted in their corresponding spatiotemporal plots
Figs.~\ref{sp_gc_t1}(b) and ~\ref{sp_gc_t1}(e). 
The globally delay-coupled network remains robust against disorder of a size as large as 
$65\%$, as shown in Figs.~\ref{poin_gc_t1}(c) and ~\ref{poin_gc_t1}(f)
for both  fixed and random values of ac torque, in which case the period-1 disorder
acquires higher order oscillations resulting in a self-organized
complex spatiotemporal representation (Figs.~\ref{sp_gc_t1}(c) 
and ~\ref{sp_gc_t1}(f)). For some distributions of $f^\prime_d$,
the network remains robust even upto $70\%$ of disorder.

\begin{figure}
\centering
\includegraphics[width=1.0\columnwidth]{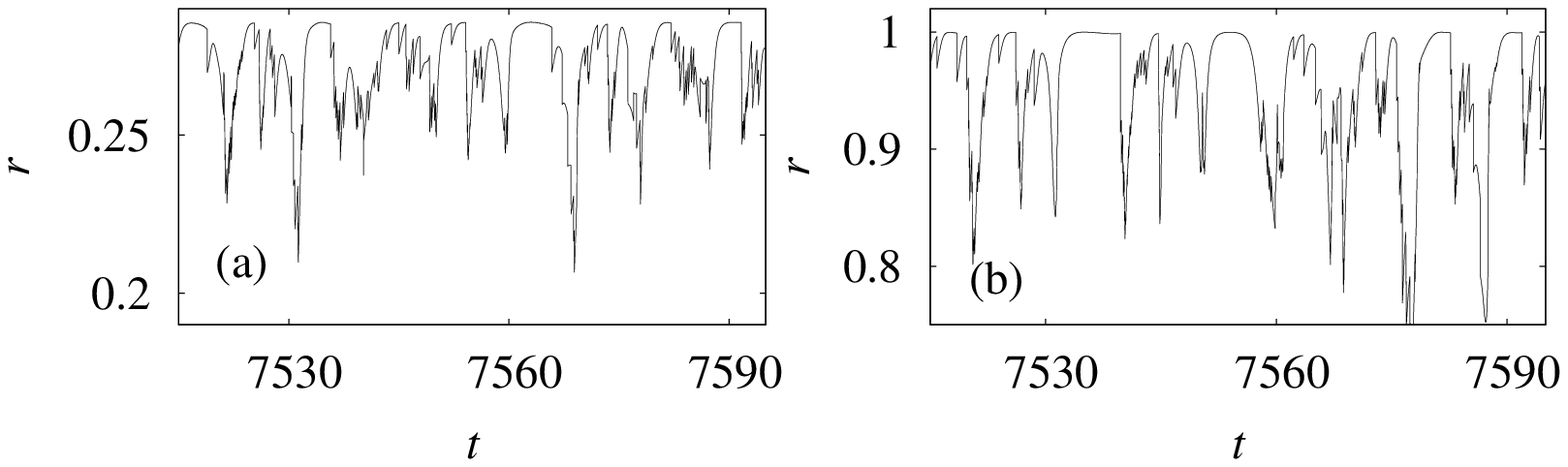}
\caption{Time evolution of the Kuramoto order parameter $r$ for
(a) $C=0.05$ and (b) $C=0.2$ with $20\%$ impurity.}
\label{global_pend_ts_op}
\end{figure}
As in the case of the ring network, we further increased the value of
delay in the coupling to examine the change in the robustness of the 
network against disorder and we obtained the same results even for
larger values of delay and for appropriate values of
coupling strength.  For instance, we present our results for
$\tau=2.0$ and $C=0.3$ in the following. 
The two largest Lyapunov exponents of the network of globally coupled pendula
and the bifurcation diagram of a pendulum in the network
are illustrated in Fig.~\ref{bif_gc}(c). We chose $f^\prime=0.96$ for a chaotic pendulum 
(as confirmed from
the positive Lyapunov exponent shown in the inset of Fig.~\ref{bif_gc}(c)) and
$f^\prime_d=1.5$ for disorders from the bifurcation diagram.
Poincar\'e points shown in Figs.~\ref{poin_gc_t2}(a) and ~\ref{poin_gc_t2}(d) 
indicate the chaotically oscillating pendula for $20\%$ 
uniform disorder for both fixed and random $f^\prime_d\in(1.2,2.0)$, respectively.
Their spatiotemporal representation is depicted in Figs.~\ref{sp_gc_t2}(a) 
and ~\ref{sp_gc_t2}(d). The evolution of the pendula in the network
in the presence of $50\%$ disorder for fixed $f^\prime_d$ is
shown in Fig.~\ref{poin_gc_t2}(b) (with its spatiotemporal plot 
in Fig.~\ref{sp_gc_t2}(b)) and for random $f^\prime_d\in(1.2,2.0)$ 
in Fig.~\ref{poin_gc_t2}(e) (with its spatiotemporal plot 
in Fig.~\ref{sp_gc_t2}(e)).  Figures~\ref{poin_gc_t2}(c) and
~\ref{poin_gc_t2}(f) exemplify the dynamical nature of the globally
coupled network in the presence of $65\%$ disorder with both fixed
and random values of ac torque. The corresponding spatiotemporal dynamics
is depicted in Figs.~\ref{sp_gc_t2}(c) and
~\ref{sp_gc_t2}(f), respectively.
The chaotic pendula remain unaltered,
while the period-1 disorders acquire higher order oscillations
for sizes larger than $50\%$ resulting in increased
spatiotemporal complexity of the original network, indicating the robustness
of the delay coupled network against disorder-induced synchronous periodic
oscillations.

Finally, as discussed in Sec.~\ref{phasecoh}, we confirm the existence of the delay enhanced 
phase-coherent oscillations in the globally connected network of pendula
 by looking at the distribution of phases in the 
$(X_i,Y_i)$  = $\left(\frac{x_{i}}{\sqrt{x_{i}^2+y_{i}^2}}, \frac{y_{i}}{\sqrt{x_{i}^2+y_{i}^2}}\right)$ plane. This is indeed shown in Fig.~\ref{gloab_cup_ps}. For $C=0.05$ (with 20\% impurity) the phases are distributed on
a large part of the unit circle (Fig.~\ref{gloab_cup_ps}(a)) and 
this reveals a poor coherence of the pendula
as confirmed by the low value of the time averaged order parameter $R=0.267$. 
The evolution of the corresponding order parameter is
depicted in Fig.~\ref{global_pend_ts_op}(a). On the other hand, for $C=0.20$ with 
$20\%, 50\%$ and $65\%$ impurities the phases are confined to a narrow region of the 
unit circle as shown in Figs.~\ref{gloab_cup_ps}(b)-(d), which is also confirmed by the 
corresponding time averaged order parameters $R=0.986, 0.988$ and $0.992$, respectively.
Also, the evolution of the order parameter $r$ corresponding to Fig.~\ref{gloab_cup_ps}(d)
is shown in Fig.~\ref{global_pend_ts_op}(b).

\section{\label{sec:level4}Summary and Conclusion}
In this paper, we have analyzed the dynamics of a regular network with 
a ring topology, and a more complex network with
all-to-all (global)  topology and studied the effect of the size of disorder.
We mainly find that the coupling delay can induce phase-coherent
chaotic oscillations in the entire network thereby enhancing the spatiotemporal
complexity even in the presence of large disorder of a size as large as $50\%$ in contrast to the
undelayed case, where even a $20\%$ disorder can render the whole network to be
periodic and thereby taming chaos. Furthermore, the delay
coupling is also capable of  increasing the robustness of the network against a large size of the disorder
upto  $70\%$ of the size of the original network, thereby
increasing the dynamical complexity of the network 
for suitable values of the coupling strength. 
We have also discussed a mechanism for the delay-induced
coherent chaotic oscillations leading to spatiotemporal complexity
in the presence of large disorders. 
We have also confirmed the delay enhanced coherent chaotic oscillations
both qualitatively and quantitatively.
We note here that the results are also robust against the size of 
the network and the size of the impurities (disorders) have to be fixed 
proportional to the size of the network.
We expect that one can use the results of our analysis to more realistic
complex networks to increase the robustness of the network against any
disorder, for example, in examining the cascading failures of complex networks,
specifically in power grids and in controlling disease spreading in epidemics, spatiotemporal
and secure communication and to increase the robustness and complexity of 
reservoir computing or liquid state machines.

\section{acknowledgments}
The work of R.~S, and M.~L is supported by the Department of Science and
Technology (DST), Government of India-Ramanna program, and also by a DST-IRPHA research project.
J. H. S is supported by a DST--FAST TRACK Young Scientist research project.
M. L. is also supported by a Department of Atomic Energy Raja Ramanna program.
D.~V.~S and J. K acknowledges the support from  EU  under project No. 240763 PHOCUS(FP7-ICT-2009-C).


\begin{thebibliography}{50}
\bibitem[{\citenamefont{Brandt et~al.}(2006)}]{expone}
\bibfnamefont{Networks with both ring and all-to-all topology exhibits
phase synchronous oscillations even with large disorders and we termed the delay 
enhanced collective chaotic oscillations exhibiting phase synchronization 
as coherent chaotic oscillations}.
\end{thebibliography}

\end{document}